\documentclass[sigconf, nonacm]{acmart}
\AtBeginDocument{%
  }

\copyrightyear{2026}
\acmYear{2026}
\setcopyright{cc}
\setcctype{by}
\acmConference[CAIN '26]{2026 IEEE/ACM 5th International Conference on AI Engineering - Software Engineering for AI}{April 12--13, 2026}{Rio de Janeiro, Brazil}
\acmBooktitle{2026 IEEE/ACM 5th International Conference on AI Engineering - Software Engineering for AI (CAIN '26), April 12--13, 2026, Rio de Janeiro, Brazil}
\acmPrice{}
\acmDOI{10.1145/3793653.3793777}
\acmISBN{979-8-4007-2475-6/2026/04}

\usepackage{framed}

\newcommand{\edit}[1]{
    \textcolor{black}{#1}
}

\newcommand{\approachName}[0]{RefineML} 

\begin{document}

\title{Applying a Requirements-Focused Agile Management Approach for Machine Learning-Enabled Systems}

\author{Lucas Romao}
\email{lromao@inf.puc-rio.br}
\orcid{0009-0001-2153-1498}
\affiliation{%
  \institution{PUC-Rio}
  \city{Rio de Janeiro}
  \state{RJ}
  \country{Brazil}
}

\author{Luíz Xavier}
\email{luiz.xavier@exa.com.br}
\orcid{0009-0008-0544-271X}
\affiliation{%
  \institution{EXA}
  \city{Parnaíba}
  \state{PI}
  \country{Brazil}
}

\author{Júlia Condé Araújo}
\email{jcaraujo@inf.puc-rio.br}
\orcid{0009-0009-8787-8062}
\affiliation{%
  \institution{PUC-Rio}
  \city{Rio de Janeiro}
  \state{RJ}
  \country{Brazil}
}

\author{Marina Condé Araújo}
\email{maraujo@inf.puc-rio.br}
\orcid{0009-0003-3374-898X}
\affiliation{%
  \institution{PUC-Rio}
  \city{Rio de Janeiro}
  \state{RJ}
  \country{Brazil}
}

\author{Ariane Rodrigues}
\email{arodrigues@inf.puc-rio.br}
\orcid{0000-0002-1614-918X}
\affiliation{%
  \institution{PUC-Rio}
  \city{Rio de Janeiro}
  \state{RJ}
  \country{Brazil}
}

\author{Marcos Kalinowski}
\email{kalinowski@inf.puc-rio.br}
\orcid{0000-0003-1445-3425}
\affiliation{%
  \institution{PUC-Rio}
  \city{Rio de Janeiro}
  \state{RJ}
  \country{Brazil}
}

\renewcommand{\shortauthors}{Romao et al.}

\begin{abstract}

Machine Learning (ML)-enabled systems challenge traditional Requirements Engineering (RE) and agile management due to data dependence, experimentation, and uncertain model behavior. Existing RE and agile practices remain poorly integrated and insufficiently tailored to these characteristics. This paper reports on the practical experience of applying \approachName, a requirements-focused approach for the continuous and agile refinement of ML-enabled systems, which integrates ML-tailored specification and agile management approaches with best practices derived from a systematic mapping study. The application context concerns an industry–academia collaboration project between PUC-Rio and EXA, a Brazilian cybersecurity company. For evaluation purposes, we applied questionnaires assessing \approachName's suitability and overall acceptance and semi-structured interviews. We applied thematic analysis to the collected qualitative data. Regarding suitability and acceptance, the results of the questionnaires indicated high perceived usefulness and intention to use. Based on the interviews, stakeholders perceived \approachName as improving communication and facilitating early feasibility assessments, as well as enabling dual-track governance of ML and software work, allowing continuous refinement of the model while evolving the overall software project. However, some limitations remain, particularly related to difficulties in operationalizing ML concerns into agile requirements and in estimating ML effort.
\end{abstract}



\received{18 November 2026}
\received[revised]{12 March 2009}
\received[accepted]{5 June 2009}

\maketitle

\section{Introduction}

Machine Learning (ML), a subset of Artificial Intelligence (AI), refers to techniques that enable systems to learn from data. The construction of ML models involves dynamic and iterative cycles of data preparation, training, testing, and model refinement, where the results are often uncertain and dependent on data quality, representativeness, and modeling choices. \cite{michalski2013machine}. These characteristics make ML development an evolving process of exploration, demanding flexible management strategies capable of accommodating experimentation and learning. The adoption of ML across industry has become a key driver of innovation. As ML matures from research prototypes to large-scale production systems, organizations increasingly recognize the need for appropriate management approaches that balance experimentation with value delivery \cite{romao2025agile}.

Agile management, a widely used iterative and adaptive approach \cite{kuhrmann2021makes}, is rooted in the principles of the Agile Manifesto \cite{fowler2001agile}. It emphasizes delivering value to customers incrementally, responding swiftly to changes, and fostering collaboration within cross-functional teams. Given its adaptability, the agile approach conceptually aligns well with the inherently dynamic and experimental nature of ML development, which often involves iterative experimentation and frequent data-driven adjustments.

However, despite these promising alignments, substantial challenges persist. Several difficulties have been reported when applying agile approaches in the ML context. For instance, teams often face unrealistic customer expectations and difficulties aligning requirements with data \cite{alves2023status}, as well as frequent misalignments between business stakeholders and ML practitioners \cite{romao2025agile, KalinowskiEtAl2025}. Additional challenges include problems in sprint planning and effort estimation due to the experimental nature of ML activities \cite{nahar2023meta, romao2025agile} and the lack of well-defined guidelines for tailoring agile management practices to ML projects \cite{vaidhyanathan2022agile4mls, romao2025agile}. Such findings indicate that, while agility provides a promising foundation for structuring ML development, its application to ML-enabled systems often requires adaptations to accommodate the characteristics of data-driven solutions and experimentation. 

Motivated by these challenges, we describe the experience and lessons learned of applying RefineML, a requirements-focused approach for continuous and agile refinement of ML-enabled systems. The experience took place within the context of a real project aimed at developing a set of ML-enabled cybersecurity features to protect users against online fraud. The project was conducted within an Industry–Academia Collaboration (IAC) between the ExACTa PUC-Rio laboratory and EXA, a Brazilian cybersecurity company that is a key player in the country’s digital protection ecosystem.

Driven by the practical need to deliver business value through continuously enhanced user protection, we defined RefineML based on the state-of-the-art academic literature. More specifically, we integrated ML-tailored requirements specification using PerSpecML \cite{villamizar2024identifying} with ML-tailored agile management using Agile4MLS \cite{vaidhyanathan2022agile4mls}, complemented by best practices identified in a recent systematic mapping study \cite{romao2025agile}.



We evaluate our experience focused on the problem-solving capability, the perceived suitability of RefineML phases and artifacts, its overall acceptance, and the identified limitations and improvement opportunities. Therefore, we applied a questionnaire, designed based on the Technology Acceptance Model (TAM) \cite{davis1989-TAM}, to the development team, and conducted semi-structured interviews with the company's business owner and the IAC project lead. We analyzed the quantitative data using descriptive statistics and the qualitative data through thematic analysis following the practical advice by ~\citet{braun2021thematic}.

Our findings indicate that RefineML was perceived as suitable for the agile management of ML-enabled system development by both customer stakeholders and the development team, addressing their main pains regarding communication, early and continuous feasibility assessment, and aligning ML and software teams. The phases and artifacts of RefineML received predominantly positive evaluations, supporting the refinement and evolution of the ML product. Its overall acceptance showed favorable perceptions of usefulness, and stakeholders expressed intent to continue using RefineML. At the same time, we identified persistent difficulties in operationalizing ML concerns into agile requirements and in estimating ML effort.

The remainder of this paper is organized as follows. Section \ref{sec:related-work} presents the background and related work. Section \ref{sec:solution-concept} introduces RefineML. In Section \ref{sec:case-study-design}, we describe the experience applying RefineML. In Section \ref{sec:results}, we present the lessons learned. Then, we discuss these lessons in Section \ref{sec:discussion}. Finally, we present the concluding remarks in Section \ref{sec:conclusion}.

  \section{Background \& Related Work}\label{sec:related-work}

\subsection{Requirements of ML-Enabled Systems}

Requirements Engineering (RE) aims to understand the problem scope and align stakeholders by discovering, documenting, and managing software requirements~\cite{sommerville1997requirements}. While it is already challenging in traditional software (SW) projects, it becomes more complex in ML-enabled systems. In the ML context, the model behaves as a learned specification derived from data \cite{villamizar2024identifying}. As a result, the system may behave incorrectly even when the code is technically correct, reinforcing the need for rigorous treatment of data requirements, acceptance criteria, and usage context. A systematic mapping study by \citet{villamizar2021requirements} shows how RE has been addressed in ML projects, describing common challenges and ML-specific requirement practices. Despite this progress, the field remains emerging and immature: validated techniques for handling ML-related non-functional requirements are still lacking, aligning model capabilities with stakeholder expectations is difficult, and empirical evidence is scarce, limiting practitioners’ ability to assess the feasibility of proposed approaches.

\begin{figure}[h!]
\centerline{\includegraphics[width=\columnwidth]{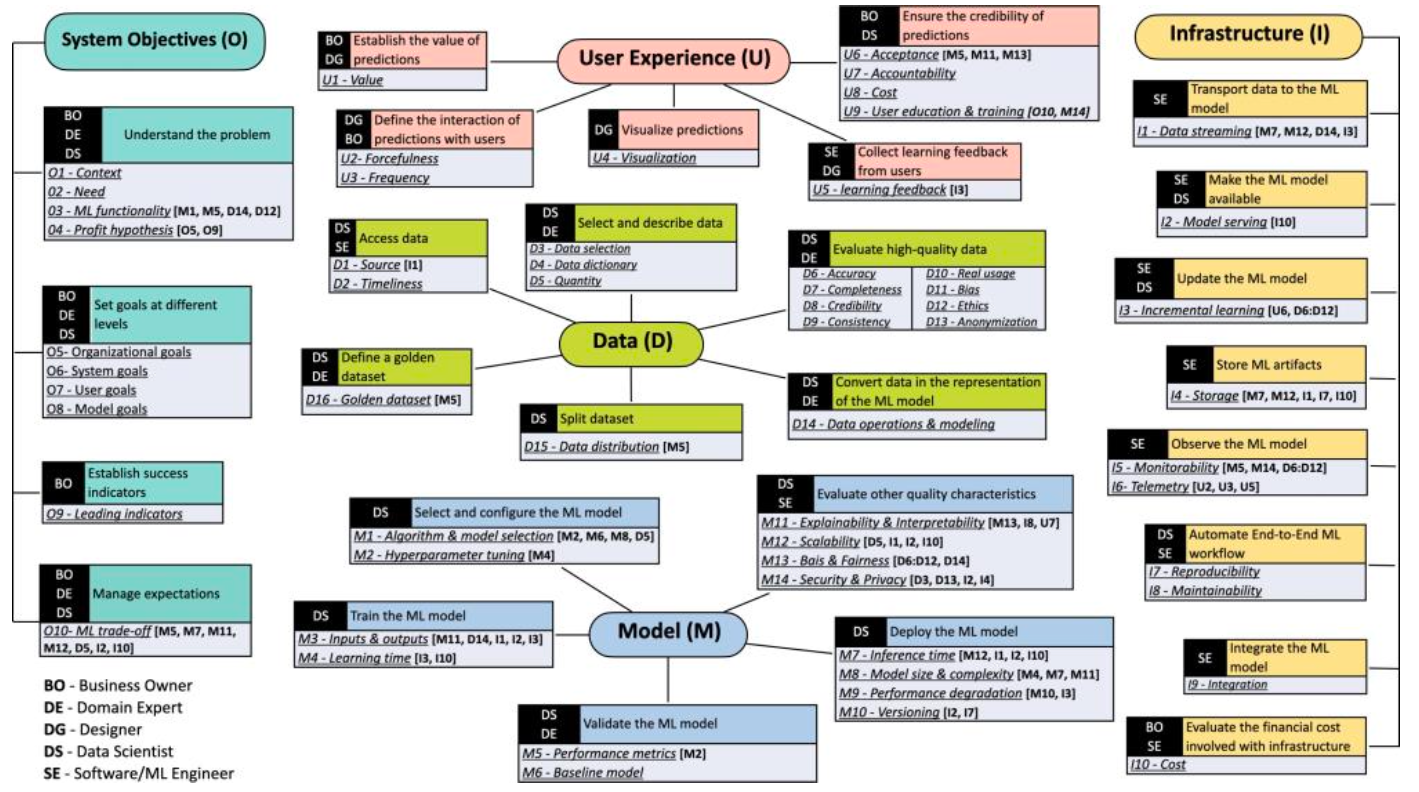}}
\caption{PerSpecML concerns illustration diagram \cite{villamizar2024identifying}.}
\label{PerSpecML}
\end{figure}

\edit{Among RE contributions for ML-enabled systems, \textit{PerSpecML} stands out as an approach built upon a solid body of empirical evidence: academic validation, industry feedback, and real-world case studies \cite{villamizar2024identifying}. It is a requirements specification approach that supports practitioners in identifying, structuring, and communicating multiple concerns involved in ML-enabled systems. The approach organizes 60 recurring concerns in ML projects into five complementary perspectives (Figure~\ref{PerSpecML}): System Objectives, User Experience, Infrastructure, Model, and Data. Since it has already been applied in real projects with positive results, demonstrating its feasibility for supporting ML system specification, we adopted it as a conceptual foundation for our work. Its five perspectives are briefly summarized hereafter.}

\textit{System Objectives Perspective} defines what the ML-enabled system aims to achieve and how success will be evaluated. It covers the functional goals, expected outputs, and performance indicators that connect business value to technical feasibility. 
This perspective ensures that ML use is justified by measurable system-level benefits and that trade-offs are made explicit.

\textit{User Experience Perspective} addresses how end users interact with and perceive the ML-enabled system. It considers factors such as explainability, trust, feedback mechanisms, ethical transparency, and model limitations. 
By including human-centric attributes, this perspective ensures that ML outcomes are interpretable and credible to diverse stakeholders.

\textit{Infrastructure Perspective} focuses on the technological environment required to support ML operations. It includes concerns such as scalability, availability, and financial cost. 
This perspective also accounts for the orchestration of pipelines, and the alignment between hardware resources and ML workload demands, ensuring operational robustness throughout the system’s lifecycle.

\textit{Model Perspective} encapsulates all aspects of ML model design, training, evaluation, and maintenance. It organizes concerns related to algorithm selection, hyperparameter optimization, performance degradation, versioning, reproducibility, and fairness. 
By treating the model as a managed artifact, this perspective promotes sustainable model lifecycle management, emphasizing traceability from requirements to evaluation metrics.

\textit{Data Perspective} structures issues surrounding data sources, quality, and governance. It includes concerns such as quantity, data preprocessing, labeling, distribution, the creation of golden datasets validated by domain experts, completeness, credibility, bias, consistency, ethics, and anonymization. 
This perspective highlights that the reliability of ML outcomes depends on data representativeness and responsible handling of sensitive information.

\subsection{Agile Management for Machine Learning}

According to a recently conducted systematic mapping study by \citet{romao2025agile}, some approaches have been proposed to align agile principles and values with the nature of ML workflows. Some examples include \textit{Agile4MLS}~\cite{vaidhyanathan2022agile4mls}, \textit{Scrum-DS}~\cite{baijens2020scrum-ds}, \textit{SKI}~\cite{saltz2019ski}, and \textit{STAMP4NLP}~\cite{kohl2021stamp}. These approaches share a common goal of bridging the gap between agile management with the sequential stages of the ML lifecycle. The mapping study identified eight key recommendations to guide the application of agile practices to ML-enabled systems. Hereafter, we provide a brief description of these recommendations.

First, \textit{Iteration Flexibility} introduces capability-based or flexible sprints to accommodate the of ML workflow. Second, \textit{ML-Specific Artifacts} such as Model, Data, and Ethical User Stories, extend traditional agile documentation to address ML-specific requirements. Third, \textit{Decoupled Ceremonies} adapt Scrum rituals to ML’s variable pace, maintaining continuous feedback without enforcing rigid sprint synchronization. Fourth, \textit{Hybrid Approaches} integrate ML workflows with agile methods. Fifth, \textit{MVMs or Demo APIs} provide minimal, functional models for integration or business validation. Sixth, \textit{Kanban Adoption} leverages Kanban’s flow-based structure. Seventh, \textit{Business Alignment} ensures ML work remains focused on stakeholder value. Finally, \textit{Ethical Considerations} safeguard against biased or undesirable model behavior.

In addition to these recommendations, the mapping study also revealed persistent challenges. The most frequently reported issue relates to \textit{sprint planning and effort estimation}, since ML tasks—such as data preprocessing, feature engineering, and model tuning—often extend beyond a single sprint and cannot be reliably scoped. Additionally, there are \textit{methodological and alignment gaps}, as organizations tend to adopt isolated agile practices without a coherent methodological foundation, resulting in miscommunication between business stakeholders, Software (SW) developers, and data scientists, which frequently leads to ambiguity in requirements and task prioritization. The last challenge involves \textit{ethical considerations}, which are rarely integrated into agile processes, despite their growing relevance in AI development.

Among the identified approaches, \textit{Agile4MLS}~\cite{vaidhyanathan2022agile4mls} stands out as an approach that facilitates the integration of agile management practices within the ML workflow. It has been successfully applied in practice and provides a structured set of practices that align agile principles with the specific characteristics of the machine learning system lifecycle. Hereafter, we describe the approach in more detail.

\begin{figure}[h!]
\centerline{\includegraphics[width=\columnwidth]{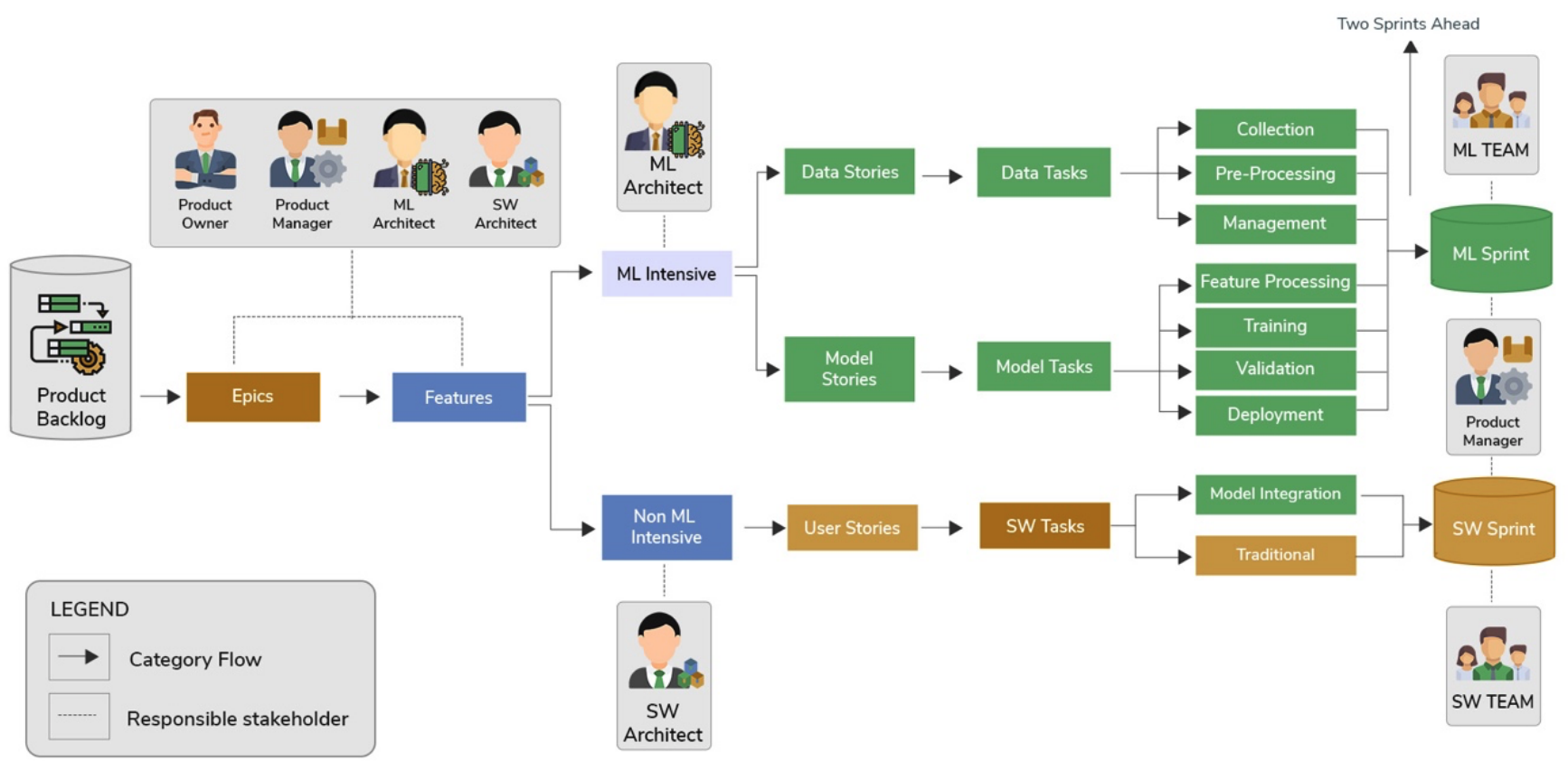}}
\caption{
The sprint planning process of Agile4MLS \cite{vaidhyanathan2022agile4mls}.}
\label{Agile4MLS}
\end{figure}

Agile4MLS is an agile management approach tailored for the development of ML-enabled systems \cite{vaidhyanathan2022agile4mls}. Its main purpose is to guide teams in systematically applying agile values, principles, and practices while accounting for the experimental, data-centric, and iterative nature of ML work. Rather than proposing a new agile framework, Agile4MLS, as presented in Figure \ref{Agile4MLS}, modifies the well-established Scrum \cite{schwaber2011scrum} cycle and artifacts, taking into account the specificities of ML projects, especially by distinguishing between backlog building for ML-intensive and non-ML-intensive features.

ML-intensive features are decomposed into two backlog items, \textit{Data Stories} (DS) and \textit{Model Stories} (MS). Data Stories mirrors well-know User Stories (US), but with a focus on the acquisition, processing, and management of datasets to ML model construction. Model Stories also mirrors US, but with a focus on feature processing, model design and training, validation, and deployment. Together, DS and MS formalize ML backlog items, promoting traceability and progress tracking across iterations.

To align the differing paces of model and product development, Agile4MLS introduces two complementary practices: the \textit{two-sprints-ahead principle} and the \textit{Demo API}. The two-sprints-ahead principle ensures that the ML team works iterations ahead of the software team, developing early prototypes and defining a Demo API, an emulated model interface that mimics the expected ML output, enabling SW developers to continue development tasks while the ML team continues model experimentation and training.

\edit{Despite the advances introduced by Agile4MLS, the problem is not solved yet. While the approach provides mechanisms to organize ML work in artifacts such as Data Stories, Model Stories, and the two-sprints-ahead principle, it does not fully address how ML-specific concerns should be identified into actionable requirements. There remains a need to explicitly specify and consider ML-related requirements from the very beginning and throughout the model refinement process.}

  \section{A Requirements-Focused Agile Management Approach for ML-enabled Systems}\label{sec:solution-concept}

\subsection{RefineML Overview}

\edit{In this section, we introduce RefineML, a requirements-focused approach for continuous and agile refinement of ML-enabled systems. PerSpecML contributes to the RE perspective, allowing us to map ML-related concerns regarding the system objectives, user experience, infrastructure, model, and data perspectives. Agile4MLS, on its part, offers a structured response to the tensions between agile practices and the ML lifecycle.}

\edit{From our previous mapping study \cite{romao2025agile}, the Minimal Valuable Model (MVM) \cite{lei2020agileclinical} stands as a concept contributing to early value delivery. Analogous to a Minimal Viable Product (MVP), an MVM represents a model that is not yet optimal but viable for deployment, capable of delivering measurable business value and validating hypotheses in real-world environments. Also, we propose a mechanism for managing model refinement through Layers of Done (LoD). Each LoD represents a specific milestone in model robustness and validation, defined by explicit evaluation metrics and acceptance criteria. This mechanism aims to allow ML teams to progressively version their models, communicate model maturity to customer stakeholders, and plan future increments in a controlled and transparent way. Within this structure, two specific layers are particularly relevant to our approach: the Demo API and the MVM. Collectively, PerSpecML, Agile4MLs, MVM and LoD practices establish the foundation for RefineML.}

\begin{figure}[h!]
\centerline{\includegraphics[width=\columnwidth]{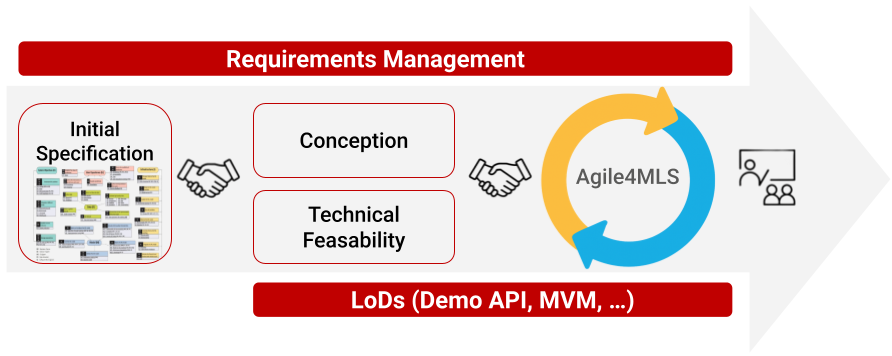}}
\caption{RefineML approach.}
\label{fig:RefineML}
\end{figure}

The approach aims to foster requirements-focused, continuous, and agile development of ML-enabled systems. To structure RefineML into phases, we borrowed the overall structure from Lean R\&D~\cite{kalinowski2025leanrd}, an approach that accelerates IAC R\&D projects through fast-paced MVP delivery, early feasibility assessment, fail-fast checkpoints, and value-focused collaboration between business and development. Lean R\&D has been successfully applied in multiple IAC projects~\cite{kalinowski2020lean, teixeira2021lessons, kalinowski2025leanrd}, including some that resulted in international patents~\cite{kuramoto2023method, kuramoto2025method}. Based on this structure, RefineML was designed to comprise the following phases: \textit{Initial Specification}, parallel \textit{Conception} and {Technical Feasibility}, and \textit{Agile Development}. Figure~\ref{fig:RefineML} illustrates these phases connecting our theoretical building blocks (PerSpecML for the initial specification, Lean R\&D's early parallel conception and technical feasibility phases and fail-fast checkpoints, Agile4MLS for agile development, continuous requirements management, and LoDs to evolve the model's Demo API into an MVM and consecutive refined versions of the model.

\subsection{RefineML activities}

\edit{In the \textit{Initial Specification} phase, the Product Owner (PO), supported by the Business Owner, Domain Expert, Designer, Data Scientist, and Software/ML Engineers, leads the specification of the ML-related requirements using PerSpecML. This early specification ensures a common understanding of what each ML component must achieve and how it will be evaluated. Importantly, the specification remains intentionally flexible, allowing the team to revise decisions as new insights emerge throughout the project.}

The \textit{Conception} phase focuses on the PO structuring two complementary backlogs, the \textit{SW Backlog} and the \textit{ML Backlog}, as proposed by \citet{vaidhyanathan2022agile4mls}. This activity evolves the PerSpecML specification into actionable agile requirements. The SW Backlog consolidates non–ML-intensive features, expressed using conventional agile requirements artifacts (\textit{e.g.}, User Stories). In contrast, the ML Backlog captures ML-intensive features, expressed through Model and Data Stories, which describe work related to data preparation, model development, evaluation, and deployment. The PerSpecML perspectives provide a structured foundation for building this ML backlog. For example, concerns such as Baseline Model (M6), Learning Time (M4), and Algorithms \& Model Selection (M1) inform the definition of Model Stories, while concerns like Source (D1) and Quantity (D5) help specify Data Stories. Bridging the two backlogs are the \textit{Enablers}, which encompass the infrastructure and architectural decisions required to support seamless integration between software components and ML pipelines. These enablers define the shared technical foundations—such as API interfaces, data pipelines, storage services, and computational resources, which will sustain both experimentation and deployment in later stages.

In the \textit{Technical Feasibility} phase, the development team (ML and SW practitioners), assisted by the Project Lead (PL), evaluates whether the ML components specified in the backlog can be realistically implemented and integrated within the project's technical and organizational constraints. This analysis is driven by three complementary assessments: \textit{Data Feasibility}, \textit{Model Feasibility}, and \textit{Demo API} construction. Data Feasibility investigates the availability, quality, and representativeness of the datasets required for training and evaluation. Model Feasibility examines the suitability of candidate algorithms and optimization techniques, considering infrastructure requirements, expected performance, and potential risks such as explainability limitations or computational costs. Finally, the Demo API is implemented to emulate the expected behavior of the ML component, enabling the software team to continue integration and interface development in parallel with ML experimentation, preserving iteration flow and enabling early validation cycles.

Together, the \textit{Conception} and \textit{Technical Feasibility} activities aim to form a bridge between strategic intent and operational execution. They ensure that the system’s software and ML dimensions can evolve coherently, that architectural and infrastructural dependencies are well understood, and that early prototypes enable agile co-development. By combining actionable specification and feasibility validation, these phases establish the foundations that sustain the iterative delivery cycles that follow.

Finally, the \textit{Development Phase} is executed following the Agile4MLS approach~\cite{vaidhyanathan2022agile4mls}, ensuring iterative delivery, continuous integration, and alignment between the business and development teams throughout the system’s evolution. Based on the outcomes of the previous phase, this stage operationalizes the SW and ML backlogs into synchronized iterations. Thanks to the Demo API implemented during Technical Feasibility, the ML team starts the phase ``two sprints ahead" of the SW team. This allows for a decoupled, yet collaborative, workflow: while the SW team focuses on implementing user stories and integrating system functionalities, the ML team iteratively refines the model, experimenting with feature engineering, algorithmic improvements, and evaluation metrics. Over time, the objective is to evolve the Demo API into a production-ready endpoint (represented by the MVM and its successive LoDs).

\edit{As represented by the red horizontal strips above and below the process flow in Figure~\ref{fig:RefineML}, requirements management is a continuous activity spanning the entire lifecycle of the project. It begins during the Initial Specification phase with PerSpecML, continues through the Conception and Technical Feasibility phase with the construction of the Backlog, and extends throughout the Development Phase via Agile4MLS’s management practices. This continuous requirements management aims at ensuring that evolving system and model requirements remain traceable and that stakeholder alignment is maintained as new insights emerge during experimentation and implementation. Furthermore, LoDs manage model evolution. The first LoD, corresponds to the Demo API created during Technical Feasibility. Another important LoD is the MVM, which is defined collaboratively between the ML team and customer stakeholders. It is noteworthy that the MVM does not necessarily represent the final expectations for the model, nor does it always immediately follow the Demo API. Similar to the MVP concept, the MVM represents a model version that can be deployed in practice to test a business hypothesis. ML practitioners may iterate in sprints to reach subsequent LoDs until reaching the layer that best satisfies business needs.}
  \section{Applying RefineML in Practice}\label{sec:case-study-design}

\subsection{Context}

\edit{RefineML was applied in the context of an ongoing R\&D Industry–Academia Collaboration project between the \textit{ExACTa PUC-Rio} laboratory and \textit{EXA}, a cybersecurity company specializing in online protection solutions. The project had the practical need to solve four interrelated challenges with the objective of increasing user protection, expanding the company's product line for online protection: (i) identify scam messages; (ii) analyze URLs to detect unsafe websites; (iii) identify screenshots that contain scam conversations; and (iv) proactively raise alerts about scams in messaging applications. This set of functions is being developed for integration into the company’s cybersecurity products.}

The development team comprises ten practitioners from the ExACTa laboratory with two to four years of experience with Software Development and ML. The team was composed of one Scrum Master, one Product Owner, three developers, and five data scientists. They were supported by a project lead (PL) from ExACTa, a PhD in computer Science with 17 years of experience with software projects, as well as a business owner (BO) from EXA with seven years of experience with software projects, and a technical lead (TL) from EXA with ten years of experience. The BO and TL jointly served as customer stakeholders throughout the collaboration. 

\edit{The cycle spanned six months, from June to November 2025. RefineML was introduced at the beginning of the cycle and used to structure all project activities. The three initial phases, Initial Specification, Conception, and Technical Feasibility, were executed within a single two-week sprint. PerSpecML supported the specification of three ML components: a scam message classifier, a malicious URL classifier, and a message extractor from screenshots. This initial specification was followed by data and model feasibility analyses, which resulted in the implementation of the message and screenshot Demo APIs. After defining the Demo API and initial definition of ML and SW backlogs, the team proceeded with Iterative Development following Agile4MLS. Throughout the project, the first author of this paper worked closely with the development team, facilitating PerSpecML workshop, guiding the introduction of the approach, and assisting in its effective application.}

\subsection{Evaluation Goal}

For evaluation purposes, our goal was to analyze the application of RefineML to characterize the approach with respect to its problem-solving capability, the contribution of each phase and its artifacts, the overall acceptance, and the identified limitations and improvement opportunities
from the point of view of the development team, project lead, and business owner in the context of an R\&D IAC project.

\subsection{Data collection}

Data collection covered perceptions from the \textit{development team} (Developers and Data Scientists), the \textit{project lead}, and \textit{customer stakeholders} involved in the project. We designed two complementary analysis instruments. A questionnaire for the development team and semi-structured interviews for the project lead and business owner. The goal was to capture both operational and managerial views on the use of RefineML.

In the development team questionnaire, we included a block of questions assessing each phase of RefineML and its artifacts. To assess the overall acceptance of the approach, we incorporated an adaptation of the Technology Acceptance Model (TAM)~\cite{davis1989-TAM}, a widely recognized model for evaluating acceptance of new technologies~\cite{turner2010does}. In this block, respondents evaluated statements related to \textit{Perceived Usefulness}, \textit{Perceived Ease of Use}, and \textit{Intention to Use} the approach in future projects. These items were answered using a five-point Likert scale.

The semi-structured interview for the project lead and customer stakeholders focused on a managerial assessment of the approach. It reused the same phases, artifacts, and core concepts from the questionnaire but presented them through items aligned with their strategic perspective. This instrument emphasized the overall suitability of the approach in supporting project objectives, handling uncertainty in ML-enabled initiatives, and fostering effective collaboration between business, software, and ML teams.

In total, we obtained twelve responses: ten from the members of the development team (through the questionnaire), one from the project lead (through an interview), and one from customer stakeholders (the BO, also through an interview). For full access to the anonymized raw data of the questionnaires and interview transcripts, please refer to the supplementary materials, which can be found in our \edit{open science repository~\cite{zenodo}}.

\subsection{Data Analysis}

We conducted quantitative analyses employing descriptive statistics for all closed-ended questions of the developer-team assessment. We analyzed the outcomes per phase of the approach, its associated artifacts, and the adapted TAM section. We computed absolute and relative frequencies, and visually presented the results using stacked bar charts to support interpretation. Responses marked as ``not applicable” were excluded from the quantitative summaries but retained to contextualize the findings in the qualitative analysis.

For the qualitative data, we analyzed the open-ended responses from the development team questionnaire and the transcripts of the semi-structured interviews with the project lead and customer stakeholders. We performed a thematic analysis procedure following the process recommended by \citet{braun2021thematic}. 



To ensure traceability, each qualitative excerpt cited in the Results is labeled using the identifier structure \textbf{XXX-YY-ZZ}, where \textit{XXX} denotes the data source (INT for interview transcripts, Q for questionnaire), \textit{YY} indicates participant role (e.g., BO for Business Owner, PL for Project Lead, DEV for Developer, DS for Data Scientist, PO for Product Owner, SM for Scrum Master), and \textit{ZZ} is a sequential index reflecting the chronological order of coded excerpts. Codes were assigned unique identifiers Cii (e.g., C01) and consistently reused across the codebook. Themes were formed through axial coding by aggregating conceptually related codes and were also assigned unique identifiers Tj (e.g., T1). The full thematic analysis materials, including all excerpts, codes, themes, and the complete audit trail, are available in our open science repository~\cite{zenodo}.

  \section{Lessons Learned}\label{sec:results}

This section presents the lessons learned from the experience of applying RefineML, integrating quantitative responses in Likert scale with qualitative data from two semi-structured interviews and open-ended questionnaire answers. We organize this section based on our evaluation goal, reporting on RefineML's problem-solving capability, suitability of the phases and artifacts, the overall acceptance, and the identified limitations and improvement opportunities.


\subsection{Problem-Solving Capability}

Over six sprints, the team implemented a robust pipeline for preventing online fraud, delivering three distinct ML-components. Although the full cycle spanned six sprints, the team achieved value delivery within the first two. The performance of the malicious message classifier MVM, deployed as a working feature, surpassed that of the preexisting solution for fraud prevention. From that point on, the team continued to improve the model in parallel with deploying the overall solution into the partner's infrastructure. 

In addition, the team delivered the complete feature for extracting messages from screenshots, including a fully functional model; only one additional sprint was required to refine this screenshot model at the Tech Lead's request. From the fourth sprint onwards, the team comprehensively restructured the data (based on Data Stories to enhance the training dataset for fraud classification) and initiated Data and Model Feasibility for the URL classification feature. Once the URL classification Demo API was ready, the team implemented the corresponding user story and dedicated the remaining time to continuous model refinement.

The final solution successfully addressed all four challenges presented by EXA through a suite of specialized ML models. The system includes components for (i) identifying scam messages and (ii) analyzing URLs to detect unsafe websites, alongside a model designed to (iii) identify screenshots containing scam conversations. Collectively, these components enable the mobile solution to (iv) proactively raise alerts regarding potential fraud. This complete defense system is now in production and fully integrated into EXA’s product line.

Regarding the qualitative analysis, the BO explicitly reported past difficulties in communicating the evolution of ML work and differentiating the product from the ML engine (INT-BO-17, INT-BO-29; C04). Stressing that ``99\%” of his role involves communication and highlighted that RefineML substantially eased this burden (INT-BO-20; C01). Furthermore, both the BO and the PL perceived that RefineML effectively supported the orchestration of the solution. Specifically, the use of dual backlogs and the Demo API reduced the risk of blocking product evolution while ML experimentation progressed (INT-BO-25, INT-PL-10, Q-PO1-03, Q-DEV1-03; C06, C10). Finally, the experience with the MVM approach ensured that a working model was available early, allowing for incremental improvements thereafter (INT-BO-26, Q-DEV3-05; C11).

\begin{framed}
\noindent
\textbf{Problem-Solving Capability Takeaway:} Overall, RefineML was suitable and able to address the customer stakeholders’ main problems by improving communication, and governance of ML work while enabling the timely delivery of functional cybersecurity capabilities.
\end{framed}


\subsection{Suitability of the Phases and Artifacts}

\begin{figure*}[ht]
\centerline{\includegraphics[width=\linewidth]{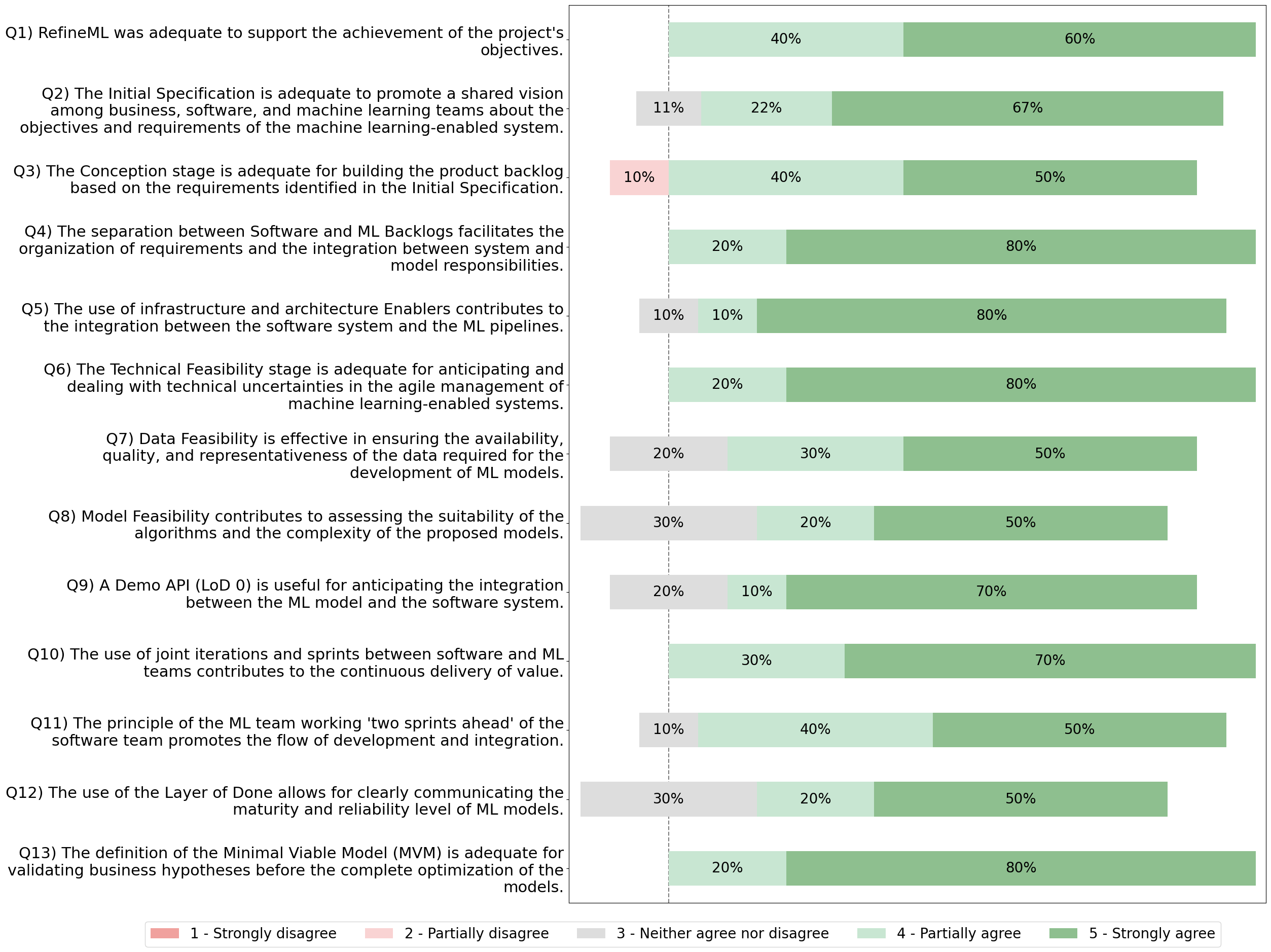}}
\caption{Suitability of RefineML phases according to the Development Team.}
\label{fig:phases_answers}
\end{figure*}

Figure~\ref{fig:phases_answers} summarizes the suitability results for RefineML phases and artifacts. While most responses reflect a positive perception, a minority of disagreeing or neutral answers highlight specific friction points, which are detailed below.

\textit{Initial Specification}: Supported by PerSpecML, this phase was widely perceived as an effective starting point for understanding each ML component. Development team reported that it successfully aligned expectations between the team and the customer (INT-PL-08, Q-PO1-01, Q-DS1-01, Q-DS2-01, Q-DS3-01, Q-DS5-01, Q-SM1-01; C02, C19). Nevertheless, one developer described the Initial Specification as ``long” and somewhat heavy for their needs (Q-DEV1-01, Q-DEV1-02; C16). To mitigate this, the Scrum Master and Product Owner noted that the phase requires strong facilitation and an active Business Owner and experienced facilitator to avoid becoming an unfocused, multi-day workshop (Q-SM1-02, Q-SM1-04, Q-PO1-02; C17, C18).

\textit{Conception}: During this phase, the dual-backlog structure effectively separated “ML experimentation” from “product delivery” (INT-PL-10, Q-PO1-03, Q-DS1-03, Q-DS4-03, Q-DS5-03; C06). Several practitioners viewed Enablers as the ``glue” between experimental ML work and concrete software integration, helping to organize the necessary infrastructure and integration tasks (INT-PL-12, INT-BO-19, Q-SM1-03; C07). Nevertheless, the concept was not straightforward for everyone; some practitioners struggled to distinguish Enablers from ordinary technical tasks or Data Stories (Q-PO1-04, Q-DEV2-04; C20), a confusion reflected in the disagreement shown in Figure \ref{fig:phases_answers}.

\textit{Technical Feasibility}: Data and Model Feasibility were consistently regarded as ``must-have” steps. The BO and PL explicitly stated that no ML project should proceed without data feasibility analysis (INT-BO-21, INT-BO-22, INT-PL-14, Q-DS5-03; C08). Similarly, Model Feasibility was valued for narrowing the search space, avoiding ``testing models forever,” and providing a structured method to compare candidate algorithms (INT-BO-24, INT-PL-16, Q-DS3-03; C09). Furthermore, the Demo API was consistently described as a ``smart move”. It defined an early contract and allowed development to proceed without waiting for a final model (INT-BO-25, INT-PL-18, Q-PO1-03, Q-DS2-03, Q-DEV1-03, Q-DEV2-03, Q-DEV3-03; C10), thereby reducing blocking dependencies.

\textit{Development using Agile4MLS}: The MVM and LoDs were central to early value delivery and incremental evolution. Both the Development team and the BO emphasized that the MVM prevented the typical ``search for 99\% accuracy” scenario by ensuring a working model was available from the start (INT-BO-26, INT-PL-20, Q-DS2-05, Q-DS4-05, Q-DEV1-05, Q-DEV3-05; C11). Concurrently, LoDs provided a transparent language to communicate model progression to stakeholders (Q-SM1-05; C23). Finally, the ``two-sprints-ahead" principle was seen as a healthy pattern for reconciling the different speeds of ML and software evolution. Data scientists reported that it gave them ``peace of mind,” affording them time to explore models while keeping pace with product development (INT-BO-27, INT-PL-22, Q-DS1-05, Q-DS5-05, Q-PO1-05; C12).

\begin{framed}
\noindent
\textbf{Suitability of the Phases and Artifacts Takeaway:} Each phase of RefineML contributed to the agile development of the ML-enabled system. 
\end{framed}


\subsection{Overall Acceptance}

\begin{figure*}[h]
\centerline{\includegraphics[width=\linewidth]{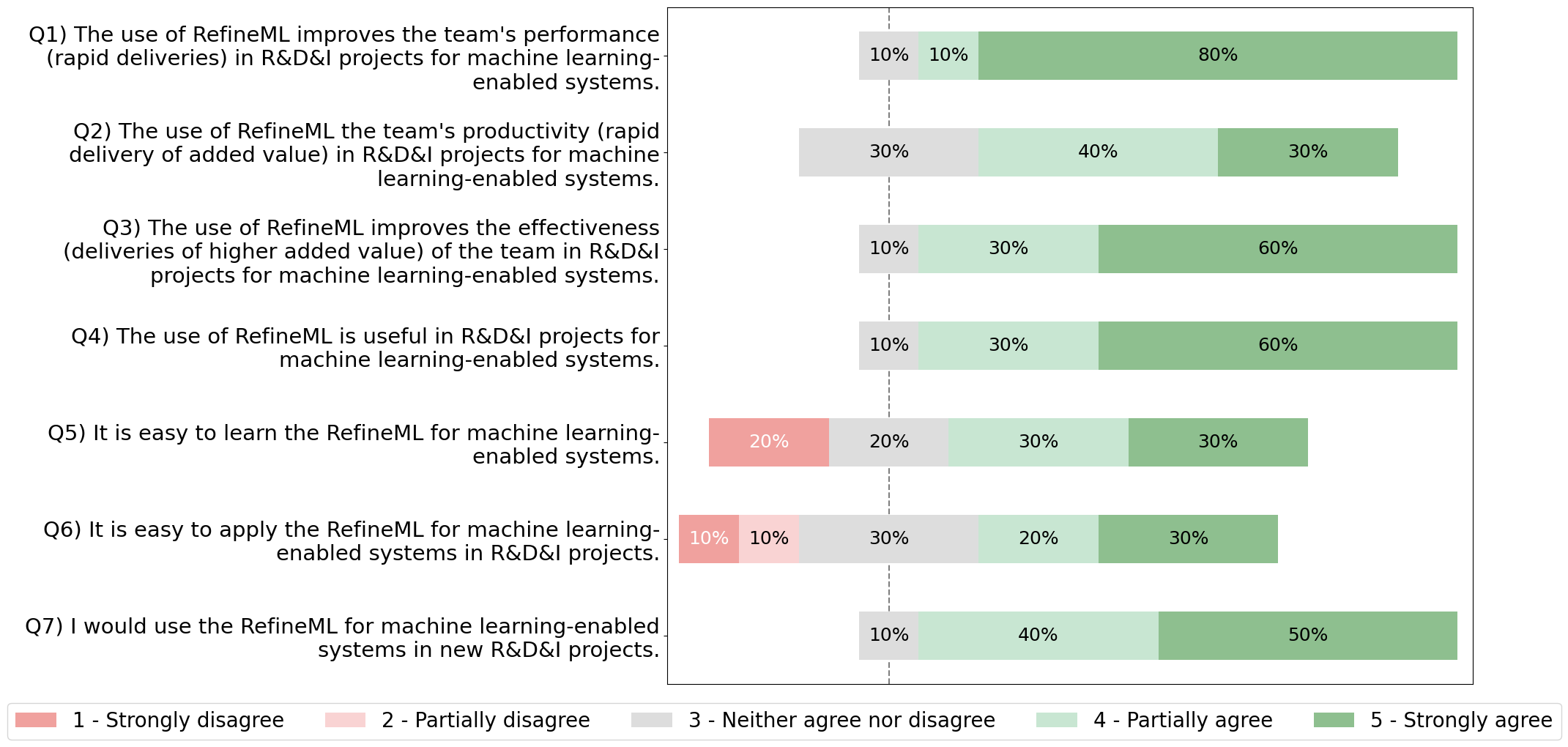}}
\caption{Technology acceptance responses from the development team.}
\label{fig:tam_answers}
\end{figure*}

Practitioners' acceptance of RefineML was assessed using TAM constructs, with Figure~\ref{fig:tam_answers} summarizing the responses. Overall, the data indicates a strong consensus that RefineML is useful for managing ML-enabled projects within an agile context.

This overall positive acceptance of all constructs is best illustrated by the BO, who explicitly stated that the approach “makes sense” and that they would likely “steal it” for their own initiatives (INT-BO-28; C13). Similarly, the Project Lead affirmed their intention to apply the method to future ML projects at ExACTa laboratory (INT-PL-24; C13).

The two questions that were most disagreed with refer to ease of learning (Q5) and ease of applying (Q6), both related to the perceived Ease of Use construct. A visible minority of respondents did not fully agree that RefineML is easy to learn or to apply, which is consistent with qualitative feedback. When assessing the characterization of these respondents, we noticed the disagreement samples shown in Figure \ref{fig:tam_answers} originate primarily from the same individuals, two junior developers (with up to two years of experience and participation in only two or three projects), who represent the least experienced members of the software team.

This reveals a divergence in perception based on experience levels. While these junior team members expressed difficulty and emphasized the need for an experienced facilitator, senior stakeholders reported the opposite. The BO explicitly described the approach as easy to use (INT-BO-29, INT-BO-30; C17), while the Project Lead characterized it as “simple, not in a diminishing way, but genuinely simple” (INT-PL-19). This contrast validates a crucial observation made by the Scrum Master and Product Owner: the approach relies on an experienced facilitator to guide the process (Q-PO1-02, Q-SM1-02; C18). It appears that without strong facilitation to bridge the gap, less experienced developers may struggle to navigate the method's artifacts, whereas seasoned professionals can more easily grasp its underlying logic

\begin{framed}
\noindent
\textbf{Overall Acceptance Takeaway:} Acceptance of RefineML was high across development team and customer stakeholders, who reported clear usefulness, intent to reuse, and willingness to recommend it. However, ease of use questions, were more divisive: junior developers noted complexity and facilitation needs, whereas more experienced practitioners described the approach as ``genuinely simple”.
\end{framed}


\subsection{Limitations and Improvement Opportunities}

Despite the benefits introduced by RefineML, a primary class of limitations concerns difficulties in operationalizing PerSpecML into ML backlog. While PerSpecML was repeatedly described as comprehensive and valuable, it was also criticized as long and demanding (Q-DEV1-01, Q-DEV1-02, Q-DS1-02, Q-DS3-02, Q-DS5-02; C16). Its success relies heavily on an engaged Business Owner and experienced facilitators capable of steering discussions and operationalizing the specification into a workable ML Backlog (Q-PO1-02, Q-SM1-02, Q-SM1-04; C17). Without this active facilitation, there is a risk of the Initial Specification becoming a ``dead document”—one that is useful for onboarding (Q-DS4-01; C02) but not revisited during execution as often as intended (INT-PL-07; Q-PO1-02; C18).

Beyond these process demands, estimation remains an unsolved challenge. Development team members reported that estimating ML effort and outcomes is difficult. The Business Owner (BO) went as far as to call unpredictability ``the only pain” they have never seen solved in any approach (INT-BO-31, INT-BO-33; C14). This sentiment was echoed by the Project Lead and other roles, who highlighted the persistent difficulty of sizing ML-intensive Stories (INT-PL-28, Q-PO1-06, Q-DS3-06, Q-DEV3-06, Q-SM1-06; C14).

Finally, ML Operations (MLOps) surfaced as persistent friction points. Even with the use of Demo APIs and Enablers, tasks such as integrating training pipelines, managing LoD versioning, and deploying models required substantial operational effort (Q-DS1-06, Q-DS2-06; C21). This friction occasionally created a sense of separation between software and ML teams (Q-DS3-04, Q-DS4-06, Q-DEV1-04, Q-DEV3-04; C22). However, it is important to note that these challenges are largely agnostic to RefineML; which suggests a need for better underlying Platform Engineering rather than changes to the approach itself.

\begin{framed}
\noindent
\textbf{Limitations and Improvement Opportunities Takeaway}: The limitations of RefineML include (i) the difficulties in operationalizing PerSpecML into ML backlog and reliance on experienced facilitation and (ii) the stochastic nature of ML, which renders accurate estimation a remaining ``unsolved pain”.
\end{framed}

  \section{Discussion}\label{sec:discussion}

The lessons learned from our experience indicate that RefineML was considered useful for managing ML-enabled systems within an agile context, while also revealing limitations that cannot be resolved by process mechanisms alone. Hereafter, we discuss the overall experience from the practitioner's perspective based on five main themes that were derived from the qualitative analysis.

\paragraph{\textbf{T1: Communication Improvement around ML work}}

The first theme, \textit{Communication Improvement around ML work} (T1), captures how RefineML strengthened alignment and expectation management by making requirements, success criteria, and constraints explicit (C01, C02, C03, C04, C05, C11, C21, C22). This structure established a shared language across business stakeholders and ML/SW practitioners, addressing prior communication pains (C04) and clarifying translation roles (C03).

The Initial Specification supported by PerSpecML revealed important to this improvement. Across development team, practitioners consistently described PerSpecML as an effective mechanism for aligning expectations and surfacing constraints early (e.g., INT-PL-08, Q-PO1-01, Q-DS1-01; C02). The Scrum Master praised it as an excellent tool for alignment and risk identification” (Q-SM1-01), while Data Scientists noted its utility for grounding the project (Q-DS2-01) and onboarding new members (Q-DS4-01). From the Business Owner's perspective—who emphasized that ``99\%” of their role is communication (INT-BO-20; C01)—RefineML directly addressed past difficulties regarding ML evolution (INT-BO-17; C04). By enabling reports to be clearly separated into product and ML updates (INT-BO-18), the RefineML, in our experience, improved visibility for stakeholders (C05).

A crucial aspect of T1 was reducing the perception of ML as magic or ``a black box.” By anchoring discussions in concrete artifacts like LoDs (C22), RefineML facilitated clearer conversations about model maturity (Q-SM1-05). The Product Owner noted a shift in customer dialogue: instead of offering vague completion percentages, they could state, ``LoD 1 is delivered, and we are working on LoD 2” (Q-SM1-05). Although some specific communication gaps remained between developers and data scientists (Q-DEV1-04, Q-DS3-04; C21), the overall perception points to a direction of the RefineML can enhance shared understanding and alignment between customer stakeholders and practitioners.

\paragraph{\textbf{T2: Feasibility Scaffolding for ML-enabled Projects}}

The second theme, \textit{Feasibility Scaffolding for ML-enabled Projects} (T2), highlights how RefineML artifacts, specifically Data Feasibility, Model Feasibility, and PerSpecML, can provide a structural framework that clarifies data requirements, technical trade-offs, and project constraints (C08, C09, C18). This scaffolding was essential for defining scope, surfacing risks early, and reducing analysis paralysis, as stated by the BO and one Data Scientist (INT-BO-26; Q-DS1-01).

Data and Model Feasibility activities were reported by both development team dn Business Owner as effective in grounding discussions in technical realities. Data Scientists noted that these practices ``saved the project” by making labeling needs and more data explicit (Q-DS3-01), as well as identifying biases that could have otherwise compromised the entire effort (Q-DS5-03). Similarly, Model Feasibility bounded the search space, preventing unfocused experimentation and endless model testing (INT-BO-24; C09). As one Data Scientist noted, the User Experience perspective revealed that ``API response time was as important as accuracy,” a constraint that effectively limited the model's scope from the start (Q-DS1-01).

Ideally, this scaffolding anchors the project in realistic expectations. The BO described Data Feasibility as a “no-brainer” without which projects are ``doomed to failure” (INT-BO-22; C08). By ensuring an upfront assessment of constraints (C18), RefineML mitigates the risk of late-stage failures due to unfeasible requirements.

Finally, the progress of the project offered insights for RefineML adaptability. The URL classifier was not prioritized until later in the lifecycle due to stakeholder decisions; yet, the team successfully applied the feasibility steps to this new component mid-stream. This suggests that RefineML can be applied not only to greenfield projects but also to the evolution of ongoing systems.

\paragraph{\textbf{T3: Dual-Track Structuring for Governing SW and ML Co-evolution}}

The third theme, \textit{Dual-Track Structuring for Governing SW-ML Co-evolution} (T3), details how RefineML’s dual-track structure orchestrates the coordination between model research and product development (C06, C07, C10, C11, C12). This structure integrates Agile4MLS principles (separate SW and ML backlogs, Demo API, and a two-sprints-ahead cadence) with practices derived from the systematic mapping (MVM and LoDs).

This governance relies on the dual backlog (C06) and the Demo API (C10) to decouple dependencies. Practitioners described the separate ML backlog as essential for making experimentation work explicit and manageable (Q-PO1-03, Q-DS1-03; C06), allowing for better organization of sprints based on model maturity (INT-PL-05). Complementing this, the Demo API was highlighted as a ``smart move” that established an early contract, thereby unblocking development work (INT-BO-25, Q-DEV1-03; C10). A developer illustrated this benefit clearly: ``I didn't have to wait for the ML team. I mocked the API (LoD 0)... and built the entire interface" (Q-DEV1-03).

To manage the differing velocities of the two tracks, the MVM (C11) and the “two-sprints-ahead” principle (C12) provided temporal buffers. Rather than waiting for a final model, teams delivered a working baseline early and improved it iteratively (Q-DS2-05, Q-DEV1-05; C11). One Data Scientist remarked that being two sprints ahead ``gives peace of mind" and prevents friction with developers (Q-DS1-05). Finally, Enablers (C07) served as the operational ``glue” (INT-BO-19, Q-SM1-03) between these tracks, ensuring that despite the separation, integration remained controlled and continuous.

\paragraph{\textbf{T4: Operationalizing PerSpecML into ML backlog}}

The fourth theme, \textit{Operationalizing PerSpecML into ML backlog} (T4), highlights the friction practitioners faced in operationalizing PerSpecML concerns into ML backlog (C15, C16, C17, C19).

A primary factor was the perceived complexity of the PerSpecML artifact (C15). Some developers found the document distant from their immediate concerns (Q-DEV1-02, Q-DS3-02), while the SM reported difficulties in decomposing its findings into concrete backlog items (Q-SM1-04; C16). This ambiguity also led to confusion regarding specific definitions, such as distinguishing between Enablers and technical tasks (Q-PO1-04, Q-DEV2-04; C19).

Consequently, the operationalization process became dependent on experienced facilitators—in this case, the first author of this paper facilitated the workshop—to steer the translation (Q-PO1-02, INT-PL-21; C16). Furthermore, if not actively maintained, this complexity increases the likelihood of the Initial Specification becoming a ``dead document” (C17). The PL and PO noted that while the specification was vital for the kickoff, it was not always treated as a living reference (INT-PL-09, Q-PO1-02), potentially leading to a disconnect between initial requirements and ongoing execution. However, this outcome was not universal. In cases where practitioners used to consult the PerSpecML specification, they reported that it proved valuable even for onboarding new members (Q-DS4-01). This contrast highlights a critical requirement: to prevent the specification from becoming obsolete, there must be incentives and mechanisms to keep the document integrated into the team's day-to-day workflow.

\paragraph{\textbf{T5: ML Effort Estimation Complexity}}

The fifth theme, \textit{ML Effort Estimation Complexity} (T5), captures the limitations stemming from ML’s intrinsic nature (C14). While RefineML seeks to manage these risks through increased structure and visibility, it still cannot eliminate the uncertainty associated with model previsibility.

Effort estimation remains a persistent challenge (C14). Practitioners highlighted the difficulty of estimating Data and Model Stories, often resorting to timeboxing as workaround (Q-SM1-06, Q-DS3-06; C14). Data Scientists pointed out that estimating outcomes, such as ``increase accuracy by 5\%”, is nearly impossible; instead, they argued it is more realistic to estimate experimentation time (Q-DS3-06). The BO explicitly framed these complexities as a remaining pain point inherent to RefineML, noting that models can sometimes degrade or show no improvement despite significant effort (INT-BO-33, INT-BO-35; C14).

Consequently, these complexities may lead to business insecurity regarding timelines and expected value (INT-BO-34). T5 underscores that because ML outcomes are iterative and stochastic, delivery predictions are inherently unreliable. While RefineML aims provides a framework to mitigate this impact (e.g., via MVM and LoDs), the core challenge of estimating the effort required to achieve specific performance targets remains an open issue.

  \section{Concluding Remarks}\label{sec:conclusion}

RefineML has been applied in an IAC innovation project between the ExACTa laboratory and EXA, a Brazilian cybersecurity company. The partnership addressed a practical need: to develop and refine ML-enabled systems that deliver business value through continuously enhanced user protection. In this paper, we reported the results of applying this requirements-focused agile approach to the continuous refinement of ML-enabled systems.

In this experience, RefineML was considered instrumental in translating complex ML requirements into impactful business results. The project delivered a comprehensive suite of cybersecurity features, including fraud detection via message and screenshot analysis, URL analysis, and proactive alerts, all of which are currently in production. Notably, the team deployed a Minimum Viable Model (MVM) early in the cycle, which already outperformed the existing solution. Hence, RefineML enabled incremental refinement without delaying value delivery, demonstrating that ML experimentation can successfully coexist with rigorous product timelines.

We also report on the perceptions of RefineML from the involved stakeholders. The assessment indicates that RefineML provided a governance structure for the co-evolution of ML and software. Stakeholders explicitly noted that the primary value lays in improved communication, visibility, and ``risk grounding" around ML work. By decoupling experimentation from product evolution via Demo APIs, the team succeeded in making previously invisible ML work explicit and manageable.

Opportunities for improvement include lowering the entry barriers for less experienced practitioners by facilitating the operationalization of PerSpecML concerns into the ML backlog (Data and Model stories). Another opportunity for evolution lies in incorporating better mechanisms for ML-complexities-aware effort estimation.

We believe that sharing this experience presents a valuable contribution to the community. RefineML was helpful in practice to facilitate communication and enable coordinated work without delaying product development. The approach was able to balance the inherent ML complexity with the predictability and early delivery required by business stakeholders. Overall, the experience highlights RefineML’s potential to support requirements-focused and agile management of ML-enabled systems in industrial contexts.

\begin{acks}

We express our gratitude to EXA for allowing us to challenge the status quo, co-creating (and trusting us to apply) innovative approaches in practice in their R\&D projects with PUC-Rio. We also thank the Brazilian Research Council - CNPq (Grant 312275/2023-4), Rio de Janeiro State's Research Agency - FAPERJ (Grant E-26/204.256/2024), the Coordination for the Improvement of Higher Education Personnel (CAPES), the Kunumi Institute, and Stone Co. for their generous support.

\end{acks}

\bibliographystyle{ACM-Reference-Format}
\bibliography{main}

\end{document}